\documentclass[conference]{IEEEtran}

\usepackage{cite}
\usepackage{graphicx}
\usepackage{amsmath,amssymb}
\usepackage{algorithm}
\usepackage{algorithmic}
\usepackage{url}
\usepackage{tabularx}
\usepackage{booktabs}
\usepackage{array}     % for \newcolumntype
\usepackage{placeins}  % for \FloatBarrier to force output of floats without forcing a new page.
\usepackage{listings}
\title{A Quick and Exact Method for Distributed Quantile Computation}

\author{
    \IEEEauthorblockN{Ivan Cao}
    \IEEEauthorblockA{
        \textit{Computer and Information Science} \\
        \textit{University of Mississippi} \\
        Oxford, MS, USA \\
        icao@go.olemiss.edu
    }
    \and
    \IEEEauthorblockN{Jaromir J. Saloni}
    \IEEEauthorblockA{
        \textit{Computer and Information Science} \\
        \textit{University of Mississippi} \\
        Oxford, MS, USA \\
        jsaloni@go.olemiss.edu
    }
    
    \and
    \IEEEauthorblockN{David A. G. Harrison\textsuperscript{*}}
    \IEEEauthorblockA{
        \textit{Computer and Information Science} \\
        \textit{University of Mississippi} \\
        Oxford, MS, USA \\
        daharri6@olemiss.edu
    }
}

\begin{document}

\maketitle
\let\thefootnote\relax\footnotetext{\textsuperscript{*}Corresponding author.}

\begin{abstract}
Quantile computation is a core primitive in large-scale data
analytics.  In Spark, practitioners typically rely on the
Greenwald–Khanna (GK) Sketch, an approximate method.  When exact quantiles are
required, the default option is an expensive global sort.  We present
\emph{GK Select}, an exact Spark algorithm that avoids full-data
shuffles and completes in a constant number of rounds. GK Select
leverages GK Sketch to identify a near-target pivot, extracts all
values within the error bound around this pivot in each partition in
linear time, and then tree-reduces the resulting candidate sets. We
show analytically that GK Select matches the executor-side time
complexity of GK Sketch while returning the exact
quantile. Empirically, GK Select achieves sketch-level latency and
outperforms Spark’s full sort by $\approx 10.5\times$ on $10^9$ values
across 120 partitions on a 30-core-node AWS EMR cluster. 
\end{abstract}

This is the extended version of the paper published in the 2025 IEEE International
Conference on Big Data (BigData), © IEEE.

\section{Introduction}\label{sec:intro}

Computing exact quantiles in large-scale distributed datasets
represents a challenge in modern data processing systems. Medians,
percentiles, and other order statistics appear frequently in
applications ranging from financial risk analysis to real-time
monitoring. In large-scale distributed systems such as Spark,
quantiles are often computed approximately to reduce overhead. While
approximate answers are sufficient in many contexts, regulatory
reporting, fairness audits, and scientific analyses often require
reproducibility or correctness guarantees that only exact quantiles
can provide.  This paper proposes an exact quantile method that
improves upon Spark's default full-sort approach.

The quantile computation problem has been extensively studied since
Hoare's seminal work on the "Find" algorithm in 1961 \cite{Hoare61},
which introduced the concept of selection without full
sorting. Subsequent theoretical advances, including Blum et al.'s
linear-time median-of-medians algorithm \cite{BFPRT72} and
Floyd-Rivest's sampling-based improvements \cite{Floyd74_select},
established strong foundations for sequential quantile
computation. However, these classical approaches face significant
scalability challenges when applied to distributed computing
environments where data is partitioned across multiple nodes and
communication costs dominate performance considerations.

Current distributed computing frameworks, particularly Apache Spark,
offer two primary approaches for quantile computation, each with
limitations. Exact quantile methods typically require a complete sort
operation, resulting in computational complexity and expensive shuffle
operations that transfer data across the entire cluster. While this
approach guarantees correctness, it becomes expensive for large
datasets. Alternatively, approximate methods such as the
Greenwald-Khanna sketch \cite{GK01} provide computational efficiency
but sacrifice exactness.

The gap between these approaches creates a need for algorithms that
can deliver exact results while maintaining computational efficiency
comparable to approximate methods. In distributed systems like Spark,
where data locality and communication minimization are critical,
traditional sequential algorithms often fail to leverage the parallel
processing capabilities effectively, resulting in suboptimal
performance that scales poorly with dataset size and cluster
resources.

This paper addresses the problem of exact quantile computation in
distributed environments by proposing GK Select, a novel algorithm
that combines the precision of exact methods with the efficiency
characteristics typically associated with approximate approaches. Our
method leverages approximate quantiles as intelligent pivot selection
for a distributed quickselect variant, reducing the computational
overhead compared to full sorting while maintaining exact results.

The remainder of this paper is organized as follows:
Section~\ref{sec:related} surveys related work in both exact and
approximate quantile computation
methods. Section~\ref{sec:implementing} describes the Spark execution
model and introduces the concepts of stage boundaries and
rounds. Section~\ref{sec:related-algorithms} describes the algorithms
compared against GK Select. Section~\ref{sec:gkselect} presents the GK
Select algorithm and its theoretical analysis. Section~\ref{Results}
evaluates performance through comprehensive experiments, and
Section~\ref{sec:conclusion} concludes.

\section{Related Work}\label{sec:related}

\subsection{Exact Selection Methods}

\subsubsection{Sequential Algorithms}
The selection problem has its roots in Hoare’s partition-based method,
where the same in-place partition primitive that powers QuickSort also
yields a linear-time, selection-by-partition scheme (FIND/QuickSelect)
in expectation \cite{Hoare61}. Worst-case linear-time selection was
later established by the median-of-medians algorithm of Blum, Floyd,
Pratt, Rivest, and Tarjan (BFPRT), which guarantees $O(n)$ time by
deterministically constructing a good pivot at each step
\cite{BFPRT72}. While BFPRT provides strong worst‑case bounds, its
constant factors often make randomized or sampling-based variants
preferable in practice. Floyd and Rivest’s SELECT refines the pivot
choice via controlled sampling to achieve expected linear time with
notably smaller constants than vanilla QuickSelect, both in analysis
\cite{Floyd73_expected} and in empirical comparisons
\cite{Floyd74_select}. These classical results clarify the trade-off
between pivot quality and iteration count that exists in nearly all
exact quantile/selection procedures.

\subsubsection{Parallel and Distributed Algorithms}

On the theoretical side, parallel selection on shared-memory PRAM
models achieves a faster time complexity. Cole introduced an optimally
efficient algorithm with linear work and $O(\log \log n)$ time on EREW
PRAM \cite{Cole88}. Han later matched the EREW lower bound with an
$O(\log n)$-time algorithm using $\frac{n}{\log n}$ processors
\cite{Han07}. Although these results are optimal within their models,
they assume unit-cost shared memory and abundant processors, which
diverges from modern data-analytics settings (e.g., Spark) where
communication, partitioning, and limited parallelism dominate
performance.

In a model closer to modern cluster computing—the Coarse-Grained
Multicomputer (CGM)—Al-Furaih et al. proposed a “serial-pivot,
parallel-count’’ selection algorithm \cite{AlFuraih97}. Their approach
uses random sampling to shrink the search interval to a small
candidate set with high probability, and then solves the reduced
problem in parallel across processors. All communication is assumed to
take place via logarithmic fan-in and fan-out, akin to Spark’s
\texttt{treeReduce} and \texttt{TorrentBroadcast}. By contrast,
Jeffers’s distributed Quickselect employs direct driver–executor
communication, but still maps naturally to Spark’s
map/shuffle/aggregate execution model and serves as a concrete
reference design \cite{Jeffers20}.

Currently, the exact quantile computation method implemented in Spark
employs a full sort. Spark’s distributed sort uses range partitioning
based on sampled keys to estimate partition boundaries (splitters),
which is structurally similar to Parallel Sort by Regular Sampling
(PSRS)~\cite{Shi92_PSRS}. Both PSRS and Spark sort shuffle a
large fraction of the records between partitions.

Spark’s sort differs from its approximate quantile computation
(\texttt{approxQuantile}) in that \texttt{approxQuantile} does not
perform a range partitioning shuffle. Instead it collects small
sketches from each executor. Because a range-partitioning shuffle
moves a large fraction of the dataset across the cluster, global
sorting is typically dominated by communication cost and is therefore
far more expensive than approximate approaches.

Practical distributed realizations of selection typically follow a
“count-and-discard’’ loop: choose a pivot, have each worker partition
locally, aggregate global counts to determine whether to search left
or right, discard the opposite side, and iterate. Open-source
implementations in message-passing environments (and in cluster
settings more broadly) embody this pattern; Jeffers’s implementation
exemplifies it, exposing an explicit round structure with pivot
broadcast, local partition/count, and global decision to “search
left’’ or “search right’’~\cite{Jeffers20}. The efficiency of such
schemes is highly sensitive to pivot quality: poor pivots increase the
number of global rounds (and thus synchronizations), whereas good
pivots collapse the search quickly. This sensitivity motivates
strategies that systematically improve pivot selection in distributed
contexts.

\subsection{Approximate Methods}

\subsubsection{Sketch-based Approaches}
Approximate quantile summaries trade exactness for strong space and
mergeability properties: attractive for streaming and distributed
data. The Greenwald-Khanna (GK) summary guarantees an approximate rank
within $\varepsilon n$ of the exact rank with worst‑case space
$\frac{1}{\epsilon}\log(\epsilon N)$\cite{GK01}. Such summaries are
per partition and combined to answer global quantile queries with
bounded rank error. Cormode\cite{Cormode17_Sketch} provides a
high-level overview of sketching. Chen et al.\cite{Chen20_Survey}
provides a comprehensive survey of the algorithmic landscape and
engineering trade-offs among deterministic sketches and randomized
summaries.

Beyond GK, a sequence of streaming sketches improves the
space--accuracy trade-off. Karnin, Lang, and Liberty’s KLL sketch
gives a mergeable randomized summary with asymptotically optimal
$O(\tfrac{1}{\varepsilon}\log\log\tfrac{1}{\varepsilon})$ space for
additive $\varepsilon n$ rank error~\cite{KLL16}. Cormode et al.'s REQ
and Masson et al.'s DDSketch instead target relative guarantees, the
former on ranks and the latter on values, and both are engineered for
distributed deployment in modern analytics and monitoring
systems~\cite{REQ21,DDSketch19}.

Furthermore, external-memory methods such as OPAQ~\cite{Alsabti97}
target one-pass approximate quantiles for disk-resident data,
recovering exact values only via a second full pass and local
sorting. OPAQ is analyzed in an I/O-centric setting and is not
instantiated or costed within a modern cluster framework like
Spark. In contrast, GK Select is a Spark-native, sketch-guided exact
algorithm that operates entirely within Spark’s RDD execution model
and comes with an explicit executor/driver complexity analysis.

\subsubsection{Sampling Methods}
Beyond deterministic sketches, sampling improves pivot or rank
estimation with minimal state, reducing communication or iteration
counts in distributed settings. In the sequential regime, SELECT’s
sampling analysis explains why better pivots lead to faster
convergence \cite{Floyd73_expected,Floyd74_select}; analogous
intuitions carry over to distributed execution where a high-quality
pivot can shrink the candidate set precipitously after one round of
local partitioning and global counting.

\paragraph*{Positioning of This Work}
Prior literature largely splits into (i) exact selection with strong
asymptotic guarantees in idealized parallel models \cite{Cole88,Han07}
or practical count‑and‑discard variants sensitive to pivot choice
\cite{Jeffers20}, and (ii) approximate quantile computation via
compact sketches with explicit rank‑error bounds
\cite{GK01,Cormode17_Sketch,Chen20_Survey}. Our approach bridges these
lines by using an \emph{approximate} summary (GK) to inform an
\emph{exact} distributed selection procedure: the sketch supplies a
pivot that is already close to the target rank, which in turn reduces
the number of global rounds and the volume of shuffled data relative
to full sorting or to pivot-agnostic distributed
Quickselect. Conceptually, this mirrors the classical role of sampling
in SELECT \cite{Floyd73_expected,Floyd74_select}, but replaces
sampling with a deterministic summary tailored to Apache Spark
\cite{GK01}.

\section{Implementing Algorithms in Spark}\label{sec:implementing}

Spark expresses computations as transformations on immutable,
partitioned datasets called \emph{Resilient Distributed Datasets
(RDDs)}. Each transformation produces a new RDD, which guarantees
determinism and fault tolerance but prevents in-place modification of
data. As a consequence, algorithms that rely on reordering keys or
performing in-place partitioning must be adapted: they create a new
dataset rather than modifying an existing one. Creating a new dataset
at least introduces a copy of the keys upon which we are computing the
$k$th order statistic. To ensure consistency and avoid recomputation,
the results of such reordering operations must be explicitly
persisted.

Spark also distinguishes between \emph{transformations} and
\emph{actions}. Transformations are lazily evaluated, whereas an
action (e.g., \texttt{collect}, \texttt{count}, \texttt{reduce})
triggers execution of the computation lineage. When an action is
invoked, Spark materializes all preceding transformations and forces
evaluation of the entire dependency chain.

We distinguish two sources of synchronization cost in Spark.

\begin{itemize}

  \item A \emph{stage boundary} arises whenever a shuffle is required:
    data must be repartitioned across executors, and no executor can
    begin the downstream stage until all upstream shuffle writes are
    complete.  Stage boundaries are a within-round cost; a single
    round may contain more than one.

  \item A \emph{round} is a unit of parallel work bounded by a driver
    synchronization barrier: executors execute in parallel, the driver
    waits for all results, and only then can it compute and broadcast
    updated information (e.g., a new pivot) to begin the next round.
    The concept of a round predates Spark and appears throughout the
    parallel computing literature: the Bulk Synchronous Parallel~(BSP)
    model formalizes computation as a sequence of \emph{supersteps},
    each consisting of a local computation phase followed by a global
    synchronization and communication step, and the Coarse-Grained
    Multicomputer~(CGM) model adopts the same structure under the name
    \emph{rounds}~\cite{AlFuraih97}.

\end{itemize}

Stage boundaries pause executors within a round; rounds impose a
coarser barrier at which the driver must act before execution can
continue. Both can dominate execution time.
   
\begin{table}[h!]
\centering
\begin{tabular}{ll}
\hline
\textbf{Symbol} & \textbf{Description} \\
\hline
$n$ & Total number of elements across all partitions. \\
$n_i$ & Total elements in the $i$th partition. \\
$\pi$ & Pivot used when performing a partition. \\
$P$ & Number of partitions. \\
$q$ & Quantile queried (e.g., $0.5$ for median). \\
$k$ & Target rank $k = n q$ \\
$r$ & The number of samples collected for splitter selection. \\
$\Delta k$ & Rank distance between approximate and exact quantile\\
\hline
\end{tabular}
\caption{Definitions used in algorithm descriptions and analysis.}
\label{tab:symbols}
\end{table}

\section{Related Algorithms}\label{sec:related-algorithms}

In this section we describe in detail each of the algorithms compared
against GK Select.  

\subsection{Spark Full Sort}

Spark computes exact quantiles by performing a global sort of the
entire dataset, implemented as a strategy similar to Parallel Sorting
by Regular Sampling (PSRS)~\cite{Shi92_PSRS}.

\begin{enumerate}
  
\item \textbf{Sampling.}
  Each executor independently selects a small,
  fixed-size sample from its local partition.
  
\item \textbf{Collect (first stage boundary)} The driver
  executes \texttt{collect} to gather all samples.  This forces
  the sampling transformation to materialize and introduces the first
  stage boundary: all executors must complete their sampling
  and send results before the driver can proceed.
  
\item \textbf{Splitter selection.} The driver sorts the samples,
  chooses $P{-}1$ splitters at target quantiles, and broadcasts them
  via \texttt{TorrentBroadcast}. The broadcast incurs communication
  from driver to executors but does not impose a stage boundary.

\item \textbf{Range partitioning (shuffle, second stage boundary).}
  Executors assign each record to one of the $P$ splitter ranges and
  emit the record to the corresponding bucket. This triggers a global
  shuffle in which every executor must send and receive data. Because
  executors cannot complete downstream tasks until the required
  shuffle data has arrived, this step forms the second stage boundary.
  
\item \textbf{Local sort.} Each executor sorts its assigned bucket
  using \texttt{UnsafeExternalSorter}. Small partitions are sorted in
  memory using dual-pivot QuickSort; large partitions spill to disk
  and are merged via multiway merge.
\end{enumerate}

\texttt{collect} and the shuffle dominate elapsed time. Both require
all executors to synchronize and transfer potentially large volumes of
data. Although the sort within each partition is efficient, the global
data movement inherent in PSRS makes Spark’s full sort a
communication-bound algorithm whose cost grows with both $n$ and the
number of partitions $P$.

\subsection{Al-Furaih Select for Spark}

We implemented a Spark-adapted variant of Al-Furaih et al.'s ``serial
pivot, parallel count'' selection algorithm~\cite{AlFuraih97}.  We
call this Al-Furaih Select~(AFS).  The method proceeds iteratively,
shrinking the search interval around the desired order statistic.

\begin{enumerate}

\item \textbf{Pivot broadcast.} Each round, the driver selects a pivot
  and broadcasts it via Spark’s \texttt{TorrentBroadcast}, a
  logarithmic-depth tree requiring $O(\log P)$ communication but no
  stage boundary.

\item \textbf{Local Partition and Count.}  \texttt{mapPartitions}
  invokes a Dutch three-way partition on each executor on each of its
  partitions, counting elements $<\pi$, $=\pi$, and $>\pi$ in their
  local partitions. Because Spark datasets are immutable, this
  produces a new RDD.  Work per partition $i$ is $O(n_i)$. Results are
  persisted for reuse in subsequent rounds.

\item \textbf{Tree reduction of counts and pivot candidates.}
  Executors send their local counts and two candidates for the next
  pivot (one below and one above the current pivot $\pi$) into a
  \texttt{treeReduce}. The reduction proceeds in $O(\log P)$ steps,
  summing counts, and discarding one candidate pivot below and one
  candidate pivot above applying reservoir sampling to maintain
  uniform pivot selection. The \texttt{treeReduce} introduces a stage
  boundary and marks the end of a round: the driver cannot proceed until
  counts from all partitions have been aggregated. \label{AFSTree}
  
\item \textbf{Driver decision and broadcast.}  The driver computes
  $\Delta k$, the rank error of $\pi$. If $\pi$ is high, it picks the
  left candidate; otherwise, the right. The new pivot is then
  broadcast via \texttt{TorrentBroadcast}, starting the next round.

\item \textbf{Iteration.} Steps 2--4 are repeated until the pivot lies
  exactly at rank $k$. The expected number of rounds is $O(\log n)$
  due to geometric shrinkage of the candidate range.
  
\end{enumerate}

Supplying candidate pivots in step~\ref{AFSTree} from both sides of
the current pivot $p$ allows the driver to choose the next pivot from
the correct direction after computing $\Delta k$ without incurring an
additional broadcast and \texttt{treeReduce} to select a pivot. This
reduces the number of \texttt{treeReduce} operations per pivot update
from two to one.

The only stage boundary in each round occurs at \texttt{treeReduce}
which happens once per round; broadcasts add latency but not
synchronization. As is revealed in Section~\ref{Results}, the rounds
imposed by the algorithm dominate runtime, while local scans
contribute only $O(n/P)$ work per executor.

\subsection{Jeffers Select for Spark}

% Corresponds to JeffersSliceQuantile.scala

Jeffers Select is identical to AFS except that Step~\ref{AFSTree}
replaces \texttt{treeReduce} with \texttt{collect}. The driver gathers
counts and candidate pivots directly from all executors, sums the
counts, and randomly selects the next pivot. Like \texttt{treeReduce},
\texttt{collect} imposes a stage boundary and marks the end of a
round. Because these messages are small and aggregation is cheap,
driver-side computation is often faster than a full
\texttt{treeReduce}, though for large $P$ the all-to-one communication
could dominate.

\begin{table}[h]
\centering
\begin{tabular}{ll}
\hline
\textbf{Symbol} & \textbf{Description} \\
\hline
$S$        & The ordered collection of tuples comprising the sketch. \\
$|S|$      & The number of tuples in the sketch. \\
$v_i$      & Sample value in strictly increasing order $v_1 < v_2 < \cdots$ \\
$g_i$      & Gap, lower bound on number of values between $v_{i-1}$ and $v_i$ \\
$\Delta_i$ & Slack, maximum rank of $v_i$ can exceed its minimum rank \\
\hline
\end{tabular}
\caption{Additional definitions used in the GK sketch description.}
\label{tab:gk-variables}
\end{table}

\subsection{GK Sketch for Spark}

If we have all the data ahead of time and can fully sort it, we could
create a summary of the data by pulling out a sample at every fifth
percentile. For example, if there are 101 data points, we could select
values at indices $0, 5, 10, 15, \ldots, 100$.  Such a summary is
approximately one fifth the size of the original data.  Given a query
value $x$, we could binary search this summary, find the closest
sample, and use its index $\times 5$ approximates the rank of $x$.
The returned rank would be guaranteed within $5\%$ of the exact rank.

Of course this strategy assumes access to the entire dataset and
requires a full sort, which is expensive in Spark if our only
objective is approximate quantile computation.

Greenwald and Khanna introduced the GK Sketch in~\cite{GK01}.  We
refer to their implementation hereafter as the \emph{Classical GK
Sketch}.  It incrementally constructs a summary while scanning the
data stream, maintaining a bounded rank error. This fits well with
Spark’s execution model, where each partition is processed as a stream
via iterators.  For example, \texttt{mapPartitions} applies a function
to a partition by providing sequential access through an iterator,
without requiring that the entire partition fit in memory.

Spark's GK sketch as of version 3.5.5, executes a variant of the
Classical GK Sketch independently across partitions and then merges
sketches back to the driver.  The final sketch can then be queried to
provide approximate ranks for any key $x$.  The Classical GK Sketch is
defined only for streams.  Spark adds an operation to merge sketches
which allows GK Sketch to be computed independently for each partition
before merging in the driver.

The Classical GK Sketch is represented as an ordered collection
of \emph{summary tuples}.

$$(v_i, g_i, \Delta_i), \quad 1 \le i \le |S|$$

where the components of this tuple are defined in Table~\ref{tab:gk-variables}.

The ordered collection may be maintained internally as a balanced tree
or similar data structure allowing $O(\log n)$ inserts.  

For every internal tuple $(v_i, g_i, \Delta_i)$ in the summary, GK Sketch
maintains the following invariant:

\begin{equation}
g_i + \Delta_{i} \le \lfloor 2\varepsilon n \rfloor \label{eq:gk_invariant}
\end{equation}

Greenwald and Khanna~\cite{GK01} prove that maintaining the Invariant
~\ref{eq:gk_invariant} guarantees that the rank of any value returned
by the summary deviates from the target rank by at most $\varepsilon
n$ in either direction.

With classical GK Sketch, when a new element $x$ arrives from the stream:

\begin{enumerate}
  \item Find its correct position in the sorted summary (via binary search on $v_i$).
  \item Insert $(x,\,g,\,\Delta)$ with $g=1$ and 
        $\Delta = g_{\text{succ}} + \Delta_{\text{succ}} - 1$ 
        (and $\Delta = 0$ at the extremes).
  \item After every $\lceil 1/(2\varepsilon)\rceil$ insertions, compress the summary by merging 
        tuples whose combined gap and slack still satisfy the invariant.
\end{enumerate}

The compress guarantees that the sketch size remains bounded by

\[S \le \frac{1}{\varepsilon} \log (\varepsilon n) + 1 \]

where $n$ is the number of elements included in the sketch so far.
This can be restated in terms of space complexity as

\begin{equation}
  S = \Theta\left(\frac{1}{\varepsilon} \log(\varepsilon n)\right) \label{eq:gk_sketch_size}
\end{equation}

Spark GK Sketch modifies Classical GK Sketch as follows:

\begin{enumerate}
  
\item Rather than inserting each arriving sample into the sketch,
  Spark GK Sketch appends the sample to the head buffer.
  
\item When the head buffer reaches maximum size $B$, the buffer is
  \emph{flushed} meaning it is sorted in $O(B \log B)$ time and then
  merged in linear time $O(B + |S|)$ into the sketch.
  
\item If the resulting sketch is larger than
  \texttt{compressThreshold}, the algorithm then compresses the sketch
  in $O(|S|)$ time in the same manner as the Classical GK Sketch.
  
\end{enumerate}

The head buffer is implemented as an array.   Appending to an array
takes $O(1)$ time.  Merging the array into the sketch takes linear time.

By performing an append onto a buffer rather than immediately
inserting each sample into the sketch, the sketch
can be implemented as a dynamic array of tuples rather than
a data structure that supports logarithmic inserts like a balanced tree.
Because the inserts only occur periodically, the merge cost
becomes amortized, but we show it does have some impact on the
time complexity analysis as Spark implements it as of this writing
and is discussed in section~\ref{GKSketchTime}.

The default value for $B$ is \texttt{defaultHeadSize = 50000}.  The default
\texttt{compressThreshold = 10000}.  Unless the buffer is forcibly
flushed before reaching $B$, flushing will also result in the sketch exceeding
the default \texttt{compressThreshold} resulting in a compress.
We thus consider the time complexity of flushing to include a call to compress.

By batching samples before compressing, Spark GK Sketch can temporarily
exceed the memory bound.  The extra tuples are pruned during the
subsequent compress after the batch insert to restore the sketch to
the $\Theta\left(\frac{1}{\varepsilon} \log(\varepsilon k)\right)$
space bound.

\subsection{Theoretical Analysis}

In this section we analyze Spark GK Sketch and modified GK Sketch.  In this analysis
we assume that the $i$th partition contains $n_i$ records and that
the data has already been distributed equally among the partitions.  As
such, for all $i$, $n_i = \tfrac{n}{P}$ where $P$ is the number of partitions.

\subsubsection{Executor Time Complexity of Spark GK Sketch}\label{GKSketchTime}

Sorting, inserting, and compressing $B$ records into
a sketch of size $S$ takes

\begin{equation}
  T_{\text{flush}} = O(B \log B + S)
  \label{eq:sgk_flush}
\end{equation}

Across $n_i$ records, we would have performed $F = \lceil n_i / B \rceil$
flushes including a final flush to handle records that had not yet
triggered a compression.

Let $T_{\text{exec}}$ denote the time complexity of the computations performed
by the executor across an entire partition.

\begin{equation}
  T_{\text{exec}} = \sum_{i=1}^F T_{\text{flush,} i}
  = \sum_{i=1}^F O(B \log B + S(k_i))
  \label{eq:sgk_T_exec1}
\end{equation}

where $S(k_i) \leq \frac{1}{\varepsilon} \log(\varepsilon k)+1$ and $k_i = B \cdot (i-1)$.
Thus Equation~(\ref{eq:sgk_T_exec1}) becomes

\[
T_{\text{exec}} = \sum_{i=1}^F
  O\left(B \log B + \frac{1}{\varepsilon} \log(\varepsilon B \cdot (i-1))\right)
\]

\begin{equation}
T_{\text{exec}} = O \left(n_i \log B + \tfrac{F}{\varepsilon} \log(\varepsilon B) 
+ \frac{1}{\varepsilon} \log\left((F-1)!\right) \right)
\label{eq:sgk_T_exec5}
\end{equation}

Stirling's formula allows us to simplify Equation~(\ref{eq:sgk_T_exec5}) to

\begin{align}
  T_{\text{exec}} &= O \left(n_i \log B + \tfrac{F}{\varepsilon} \log(\varepsilon B) 
  + \frac{1}{\varepsilon} F \log F \right) \notag\\
  &= O \left(n_i \log B + \frac{1}{\varepsilon} \tfrac{n_i}{B} \log(\varepsilon B) 
    + \frac{1}{\varepsilon} \tfrac{n_i}{B} \log \tfrac{n_i}{B} \right) \notag \\
  &= O \left(n_i \log B + \frac{1}{\varepsilon} \tfrac{n_i}{B} \log(\varepsilon n_i) \right)
    \label{eq:sgk_T_exec7}
\end{align}

Since we assume $n_i = \tfrac{n}{P}$,

Equation~(\ref{eq:sgk_T_exec7}) becomes

\[
T_{\text{exec}}  
    = O \left(\tfrac{1}{\varepsilon} \tfrac{n}{P} \log(\varepsilon \tfrac{n}{P})
    \right)
\]

However, this is only valid if 
$\tfrac{1}{\varepsilon} \tfrac{\log(\varepsilon n_i)}{B} \gg \log B$.  With Spark’s
defaults ($\varepsilon \approx 10^{-2}, B=5 \times 10^4$) the inequality is false for
any achievable dataset.

Assuming base 2 logarithms,

\[\tfrac{1}{\varepsilon} \tfrac{\log(\varepsilon n_i)}{B}\gg \log B\]

\[
  n_i \gg \tfrac{1}{.01} (5 \times 10^4)^{500}
\]

which is vastly more than the number of atoms in the universe $(\approx 10^{80})$, so we should not
eliminate the first term.

\[
T_{\text{exec}} = O \left(\frac{n}{P} \log B
  + \frac{1}{\varepsilon} \tfrac{n}{P}\tfrac{1}{B} \log (\varepsilon \tfrac{n}{P})\right)
\]

This is not the same time complexity as derived for the Classical GK Sketch.

\subsubsection{Driver Time Complexity of Spark GK Sketch}

With Spark's \texttt{approxQuantile} (as of current versions, including Spark
3.x and later), the GK driver then performs a \texttt{collect} action to
transfer the per-partition sketches from the executors back to the
driver. On the driver, it merges them sequentially via a \texttt{foldLeft},
i.e., an iterative pairwise merging.  This forms the global sketch
before querying the final approximate quantiles.

Due to space constraints, we omit the derivations of the driver
time complexity, but we point out that \texttt{foldLeft}
results in asymptotically worse performance than if the driver
were to instead locally perform a recursive tree reduce.

With \texttt{foldLeft}, merging dominates yielding

\begin{equation}
T_{\text{driver}} = \Theta\!\left(
      \frac{P}{\varepsilon}\log(\varepsilon n) 
      \right)
\end{equation}

\subsubsection{Modified GK Sketch}

To recover the asymptotic behavior of the classical GK sketch, we consider a
lightweight modification of Spark’s implementation.  The goal is to avoid
fixed-size buffering effects and the linear \texttt{foldLeft} merge on the
driver.  The changes are

\begin{enumerate}
  
  \item{$B$ starts small.  After each flush+compress, set
    $B \leftarrow \lceil \alpha\,|S|\rceil$, where $\alpha > 1.$}
    
  \item{Sketches tree reduce recursively in the driver.}
    
\end{enumerate}

Note that we have not modified Spark.  We use the stock implementation
Spark GK Sketch for performance results in Section~\ref{Results}.
This modified version is used in analysis only.

Let $T_{\text{mSGK}}$ denote the per-insertion time complexity for the Modified
Spark GK Sketch (mSGK) implementation.

\[
T_{\text{mSGK}} = T_{\text{insert}} + T_{\text{compress}}
\]

\begin{align}
  T_\text{compress} &= \frac{1}{n} \sum_{j=1}^k O(\alpha |S(n_j)|) \notag \\
  &=  \frac{1}{n} O\!\left(\alpha |S(n_1)| +
      \alpha |S(n_2)| + \cdots + \alpha |S(n_k)|\right)
\label{eq:msgk_k}
\end{align}

where $k$ is the number of calls to compress and $j$ is the $j$th call to compress.
The $\tfrac1n$ factor appears because we are amortizing the cost of the compressions
across $n$ inserts.

\medskip
Let us use a continuous approximation.

\[
H(n) = \alpha | S(n)| = \frac{\alpha}{\varepsilon} \log(\varepsilon n)
\]

\[
k \approx \int_{n_0}^n \frac{dx}{H(x)} 
    = \tfrac{\varepsilon}{\alpha} \int_{n_0}^n \frac{dx}{\log(\varepsilon x)}
\]

Let $u = \varepsilon x$ and $du = \varepsilon dx$, so the above becomes

\[
k \approx \tfrac{1}{\alpha} \int_{\varepsilon n_0}^{\varepsilon n} \frac{du}{\log u}
= \tfrac{1}{\alpha} \left( \operatorname{li} (\varepsilon n) - \operatorname{li} ( \varepsilon n_0 ) \right)
\]

\[
\operatorname{li}(x) = \frac{x}{\log x} + O\!\left(\frac{x}{(\log x)^2}\right) \approx \frac{x}{\log x}
\]

\[
k \approx \tfrac{1}{\alpha} \left( \operatorname{li} (\varepsilon n) - \operatorname{li} ( \varepsilon n_0 ) \right)
    = \frac{1}{\alpha} \frac{\varepsilon n}{\log (\varepsilon n)}
    - \frac{1}{\alpha} \frac{\varepsilon n_0}{\log (\varepsilon n_0)}
\]

Assuming $n \gg n_0$, we can rewrite the above as 

\begin{equation}
k \approx \frac{1}{\alpha} \frac{\varepsilon n}{\log (\varepsilon n)}
\label{eq:msgk_k2}
\end{equation}

Amortizing over $n$ insertions,

\begin{equation}
T_{\text{compress}} = \tfrac{1}{n} \sum_{j=1}^k O(\alpha |S(n_j)|) 
                         < O\!\left(\tfrac{k}{n} \cdot \alpha \cdot \tfrac{1}{\varepsilon} \log(\varepsilon n)\right)  \label{eq:msgk_T_comp}
\end{equation}

Substituting \eqref{eq:msgk_k2} into \eqref{eq:msgk_T_comp} yields the
amortized compress time 

\[
T_{\text{compress}} = O\!\left(\frac{1}{n} \cdot \frac{1}{\alpha} \frac{\varepsilon n}{\log (\varepsilon n)} 
                  \cdot \alpha \cdot \frac{1}{\varepsilon} \log(\varepsilon n)\right)
\]

which simplifies to

\[
T_{\text{compress}} = O(1)
\]

Appending to the buffer takes $O(1)$ until the buffer is full. Because
copying into a new buffer is $O(B)$ where $B_j = \alpha \cdot |S_{j-1}|$, an
analogous derivation as was used for compressing yields $O(n)$
cost for $n$ appends, giving amortized $O(1)$ per append.

However, before a buffer can be merged into the sketch it must be ordered,
which takes $B_j \log B_j$ time. Thus

\begin{align}
T_{\text{insert}} &= O(1) + \frac{1}{n} \sum_{j=1}^k O(B_j \log B_j) \notag \\
   &= O(1) + \frac{1}{n} \sum_{j=1}^k O( |S_j| \log |S_j| )
\label{eq:msgk_T_exec1}
\end{align}

where

\[
\sum_{j=1}^k O( |S(n_j)| \log |S(n_j)| ) <  O\!\left(k \cdot |S(n)| \cdot \log |S(n)|\right)
\]

so Equation~\eqref{eq:msgk_T_exec1} becomes

\begin{equation}
T_{\text{insert}} = O\!\left( \frac{1}{n} \cdot k \cdot |S(n)| \cdot \log|S(n)|\right)
\label{eq:msgk_T_exec2}
\end{equation}

Substituting Equations~\eqref{eq:msgk_k2} and \eqref{eq:gk_sketch_size}
into \eqref{eq:msgk_T_exec2} yields

\[
T_{\text{insert}} = O\!\left(  
  \frac{1}{n} \cdot \frac{1}{\alpha} \cdot \frac{\varepsilon  n}{\log (\varepsilon n)} 
  \cdot \frac{1}{\varepsilon} \log(\varepsilon n) \cdot \log\!\left(\tfrac{1}{\varepsilon} \log(\varepsilon n)\right)\right)
\]

which simplifies to

\begin{equation}
T_{\text{insert}} = O\!\left( 
  \frac{1}{\alpha} \log\!\left(\tfrac{1}{\varepsilon} \log(\varepsilon n)\right)\right)
\label{eq:msgk_T_exec4}
\end{equation}

Because $T_{\text{compress}} = O(1)$, the above becomes

\begin{equation}
  T_{\text{mSGK}} = O\!\left( \log \tfrac{1}{\varepsilon} + \log \log(\varepsilon n)\right)
  \label{eq:msgk_T_exec5}
\end{equation}

Often the first term is viewed as a constant, causing Equation~\eqref{eq:msgk_T_exec5} to simplify to

\[
T_{\text{mSGK}} = O\!\left( \log \log(\varepsilon n)\right)
\]

This is the same time complexity as the Classical GK Sketch.  However,
because we may later tune $\varepsilon$ to control the tradeoffs
within GK Select, we prefer to keep the form presented in
Equation~\eqref{eq:msgk_T_exec5}.

When we insert all elements of a partition of size $n_i = \tfrac{n}{P}$,
Equation~\eqref{eq:msgk_T_exec5} becomes

\begin{equation}
T_{\text{exec}} = O\!\left(\tfrac{n}{P}\log \tfrac{1}{\varepsilon} 
                  + \tfrac{n}{P} \log \log\!\bigl(\varepsilon \tfrac{n}{P}\bigr)\right) 
\label{eq:msgk_T_exec}
\end{equation}

For large $n$ and small $\varepsilon$ the above becomes

\begin{equation}
T_{\text{exec}} = O\!\left(\tfrac{n}{P} \log \log\!\bigl(\varepsilon \tfrac{n}{P}\bigr)\right)
\end{equation}

Because we scale $B$ as a constant factor larger than the space bound,

\begin{equation}
S_{\text{exec}} = O\!\left(\tfrac{1}{\varepsilon} \log(\varepsilon \tfrac{n}{P})\right) 
\label{eq:msgk_S_exec}
\end{equation}

Rather than \texttt{foldLeft}, if the driver implements a
\texttt{treeReduce} that executes entirely in the driver, the time
complexity becomes:

$$T_{\text{driver}} = \Theta\left( 
        P \cdot \tfrac{1}{\varepsilon} \log(\varepsilon \tfrac{n}{P})\right)
   \label{eq:msgk_T_d}$$

The driver and executor space complexity, network volume, and network latency
remain unchanged from Spark's GK implementation.

\section{GK Select Algorithm}\label{sec:gkselect}

To compute the $k$-th smallest element efficiently under the Spark
execution model, we propose \emph{GK Select}, an exact quantile
algorithm guided by an approximate pivot.

\begin{enumerate}

\item \textbf{Executors compute local GK sketches.} 
  The first phase of \texttt{approxQuantile} runs on each executor,
  building a GK sketch for its local partition. This corresponds to the
  executor-side work of Spark’s GK implementation.

\item \textbf{Driver collects and merges GK sketches}
  After the executor-side sketches are completed, \texttt{approxQuantile}
  issues a \texttt{collect} to gather all sketches at the driver, which
  merges them to produce a global sketch and compute an approximate
  quantile.

\item \textbf{Driver broadcasts approximate pivot.}  The driver then
  broadcasts the approximate quantile to
  all executors using \texttt{TorrentBroadcast}.  The approximate quantile
  is used in subsequent steps as the pivot $\pi$.
  
\item \textbf{Executors count.} Within \texttt{mapPartitions},
  each executor counts $C_i$ the number of elements $<\pi$ in
  each of its local partitions.
  
\item \textbf{Driver collects counts.}
  The driver executes \texttt{collect} to gather all $C_i$ and computes
  $\Delta k = k - \sum_i C_i$, the distance between the approximate and
  exact ranks. If $\Delta k = 0$, the pivot is exact.  Done.

\item \textbf{Driver broadcasts correction.}
  The driver broadcasts $\Delta k$ to all executors, which
  is signed so it conveys whether $\pi$ was above or below $k$.
  
\item \textbf{Executor selects $\Delta k$ candidates for exact quantile}  Each executor runs
  a Dutch partition within \texttt{mapPartitions} around
  $\pi$.  If $\pi < k$, each executor uses QuickSelect to extract the 
  $\Delta k$ smallest values above $\pi$; 
  if $\pi > k$, it extracts the $|\Delta k|$ largest below $\pi$.
  The exact quantile will be in one of the ranges from the
  partitions.

\item \textbf{Reduce across executors.} 
  The extracted candidate sets are merged via \texttt{treeReduce}. 
  At each step, only values within $\Delta k$ rank of $\pi$ are retained; 
  others are discarded.

\item \textbf{Driver finds the exact quantile.} 
  After the tree reduce at the driver, $\Delta k$ candidates remain, 
  representing the gap between the approximate and exact quantile. 
  The boundary value (or its neighbor if parity requires) is the exact quantile.

\end{enumerate}

A variant of Step~2 could run a Dutch partition immediately and
infer counts from the pivot position, then reuse the partitioned RDD
during candidate extraction in the local partitioning step. However,
this would require persisting a full intermediate dataset. By instead
performing a Dutch partition around $\pi$ and then QuickSelect to the
$\Delta k$ closest elements (in rank) to $\pi$ during the
local-partitioning step, we avoid any persists while maintaining
linear time.

\subsubsection{GK Select Executor Time Complexity}

\begin{table}[h]
\centering
\begin{tabular}{|c|c|l|}
\hline
\textbf{Time Complexity} & \textbf{Step} \\
\hline
$T_{\text{GK,e}}$           & 1   \\
$T_{\text{GK,d}}$           & 2   \\
$T_{\text{b,d}}$            & 3   \\
$T_{k,\text{e}}$            & 4   \\
$T_{\text{collect,d}}$       & 5   \\
$T_{\text{bk,d}}$           & 6   \\
$T_{\text{select,e}}$        & 7   \\
$T_{\text{reduce,e}}$        & 8   \\
$T_{\text{final,d}}$         & 9   \\
\hline
\end{tabular}
\caption{Time complexity notation for GK Select.  Substitute $S$ for $T$
to denote the analogous space complexity.}
\label{tab:gk-complexity}
\end{table}

\begin{equation}
T_{\text{exec}} = T_{\text{GK,e}} + T_{k,\text{e}} 
    + T_{\text{select,e}} + T_{\text{reduce,e}}
\end{equation}

For purposes of this analysis we adopt the Modified Spark GK Sketch
implementation, because it preserves the time complexity of Classical
GK Sketch.

\begin{equation}
T_{\text{GK,e}} = O\!\left(\tfrac{n}{P}\log \tfrac{1}{\varepsilon} 
    + \tfrac{n}{P} \log \log(\varepsilon \tfrac{n}{P})\right)
\end{equation}

Counting the number of records below the pivot takes linear time:

\[
T_{k,\text{e}} = O\!\left(\tfrac{n}{P}\right)
\]

Candidate selection in Step~7 performs a local Dutch partition around
the pivot and then a QuickSelect to find the $\Delta k$ relative to
the pivot, yielding

\[
T_{\text{select,e}} = O\!\left(\tfrac{n}{P}\right)
\]

During a tree reduce there are $P$ leaves, resulting in $\Theta(\log
P)$ tree depth\footnote{The conference version ignored the $\log P$
factor in the tree-reduce span.  Under the regime $P \le
1/\varepsilon$, the corrected term $\varepsilon n \log P$ is still
asymptotically smaller than $\frac{n}{P}\log(1/\varepsilon)$ and thus
drops out; the simplified time complexity reported in Table~III of the
conference version remains unchanged.} in Step~7.  Each merge handles
at most $\Delta k\le \varepsilon n$ elements, so the time complexity
becomes

\[
T_{\text{reduce,e}} = O(\varepsilon n \log P)
\]

Summing terms yields

\[
\begin{aligned}
T_{\text{exec}} &= O\!\left(\tfrac{n}{P}\log \tfrac{1}{\varepsilon} 
    + \tfrac{n}{P} \log \log(\varepsilon \tfrac{n}{P})\right) \\
    &\quad + O\!\left(\tfrac{n}{P}\right) + O\!\left(\tfrac{n}{P}\right) + O(\varepsilon n \log P)
\end{aligned}
\]

which simplifies to

\[
T_{\text{exec}} = O\!\left(\tfrac{n}{P}\log \tfrac{1}{\varepsilon} 
    + \tfrac{n}{P} \log \log(\varepsilon \tfrac{n}{P}) + \varepsilon n \log P \right)
    \]

For cluster sizes $P \leq \tfrac{1}{\varepsilon}$, this simplifies to

\[
T_{\text{exec}} = O\!\left(\tfrac{n}{P}\log \tfrac{1}{\varepsilon} 
    + \tfrac{n}{P} \log \log(\varepsilon \tfrac{n}{P}) \right)
\]

In interior nodes of the tree reduce, the working set may be as large
as $O(\varepsilon n)$, which can exceed $\tfrac{n}{P}$.  With
extremely large clusters, the assumption that $O(\varepsilon n) <
O(\tfrac{n}{P})$ breaks down.  Given that the last merge occurs on the
driver and drivers are often less endowed than executor nodes, this
could cause the driver to become a bottleneck.
    
\subsubsection{GK Select Driver Time Complexity}

\begin{equation*}
T_{\text{driver}} = T_{\text{GK,d}} + T_{\text{b,d}} + T_{\text{collect,d}} + T_{\text{bk,d}} + T_{\text{final,d}}
\end{equation*}

Cost to collect and merge GK sketches using the Modified Spark GK Sketch implementation yields:
\begin{equation*}
T_{\text{GK,d}} = \Theta\!\left(\tfrac{P}{\varepsilon}\log(\varepsilon \tfrac{n}{P})\right)
\end{equation*}

Torrent broadcast to distribute the approximate quantile:
\begin{equation*}
T_{\text{b,d}} = O(1)
\end{equation*}

Cost to collect counts, sum them, and determine the difference between the approximate and the exact quantile:
\begin{equation*}
T_{\text{collect,d}} = O(P)
\end{equation*}

Torrent broadcast of the difference between the approximate and the exact quantile:
\begin{equation*}
T_{\text{bk,d}} = O(1)
\end{equation*}

Time to find the minimum or maximum of the numbers returned from the \texttt{treeReduce} of candidates:
\begin{equation*}
T_{\text{final,d}} = O(\varepsilon n)
\end{equation*}

Summing components yields

\[
T_{\text{driver}} = O\!\left(\tfrac{P}{\varepsilon}\log(\varepsilon \tfrac{n}{P}) 
                    + \varepsilon n\right)
\]

\subsubsection{GK Select Executor Space Complexity}

GK Select incurs the space requirements of the Classical GK Sketch on
each executor. In addition, the Dutch partition and QuickSelect steps
operate on a fully materialized view of the partition inside
\texttt{mapPartitions}. Under Spark’s iterator model, this means that,
for each partition, the executor must hold a single in-memory copy of
the partition while performing the partitioning and candidate
extraction; we do not invoke \texttt{persist}, and no second copy of
the RDD is created by the algorithm itself.

As a result, the peak executor space usage is

\[
S_{\text{exec}} = O(\tfrac{1}{\varepsilon} \log(\varepsilon \tfrac{n}{P}) 
+ \tfrac{n}{P})
\]

Because a sketch can never contain more than all of the keys in the
partition, the second term dominates yielding

\[
S_{\text{exec}} = O(\tfrac{n}{P})
\]

\subsubsection{GK Select Driver Space Complexity}

To implement GK Select, the driver incurs all of the space requirements
of GK Sketch plus the space required to hold the final $\Delta k$ closest
candidates to the approximate quantile,

\[
S_{\text{driver}} = O\left(\tfrac{P}{\varepsilon}\log(\varepsilon \tfrac{n}{P}) + \varepsilon n\right)
\]

\subsubsection{GK Select Network Volume Across Cluster}

Several exchanges of messages take place.

\begin{itemize}

\item \texttt{collect} sketches to the driver.

  $$O(\tfrac{P}{\varepsilon} \log(\varepsilon \tfrac{n}{P}))$$
   
\item \texttt{TorrentBroadcast} approximate quantile from the driver to the executors. \

  $$O(P)$$

\item \texttt{collect} counts of the number elements lesser than the approximate.

  $$O(P)$$

\item \texttt{TorrentBroadcast} $\Delta k$ the difference between the approximate and exact quantile.

  $$O(P)$$

\item \texttt{treeReduce} the $\Delta k$ candidates from each executor to the driver where $|\Delta k| \le \varepsilon n$.

  $$O(\varepsilon n P)$$
\end{itemize}

Network volume sums to

$$O(\tfrac{P}{\varepsilon} \log(\varepsilon \tfrac{n}{P})) + O(P)+ O(\varepsilon n P)$$

which simplifies to

$$O(\tfrac{P}{\varepsilon} \log(\varepsilon \tfrac{n}{P}) + \varepsilon n P)$$

\subsubsection{Algorithm Tradeoffs}

We assemble time complexities for all algorithms in
Tables~\ref{tab:comparison-time-mem} and~\ref{tab:comparison-comm}.
Asymptotically AFS and Jeffers Select clearly have the best executor
time complexity.  Assuming that the values are evenly distributed
across the partitions we would expect near linear speed up.  However a
significant difference in cost structure appears in
Table~\ref{tab:comparison-comm}: AFS and Jeffers Select both update a
pivot such that it converges to the exact $k$th order statistics in
$O(\log n)$ rounds.  Every round imposes a driver synchronization
barrier preventing progress until the next pivot value is determined.

AFS appears to have better driver time complexity, but this is
achieved by introducing a \texttt{treeReduce} of the counts rather
than a \texttt{collect}.  A \texttt{treeReduce} will always have more
communication overhead to set up the tree spanning executors, which it
recovers if the computation at each reduce step is substantial, but in
the case of AFS the only computation being performed at each node in
the reduce is to sum the counts of the number of values below the
pivot and select one candidate pivot from below and one candidate
pivot from above the current candidate $\pi$.  Only in a regime with
extremely large clusters might AFS outperform Jeffers Select.

GK Select has the same executor time complexity as the classical GK
Sketch, but this is only because a Dutch partition and Quickselect
run in linear time.  The additional operations performed by GK Select
will increase the constant factor hidden by time complexity analysis
relative to GK Sketch.

The biggest downside of GK Select appears to be on the driver side.
All candidates within $\Delta k$ rank of the approximate quantile are
reported to the driver and then driver selects the exact quantile
as the farther of the candidates.  This introduces additional
driver side time and space complexity of $O(\Delta k)$
where $\Delta k \le \varepsilon n$.  If $\varepsilon > \tfrac{1}{P}$
then the number of keys processed by the driver can exceed the
keys in the average-sized partition.  Because driver nodes are often
less well-endowed than executors this could become a problem.

If $\varepsilon$ is set too low then the size of the sketches
may limit performance given that sketches grow roughly
inversely proportional to $\varepsilon$.  One might be tempted to
set $\varepsilon = \tfrac{1}{P}$, but then driver time and space
complexity grows roughly with the square of the number of partitions.
If we intend to process such large GK sketches, it might make
sense to perform a \texttt{treeReduce} when merging sketches
between partitions rather than performing a \texttt{collect} and
merging on the driver.  Doing so would push merge overhead into the
cluster allowing us to accommodate much larger sketches,
achieving proportionally lower error in \texttt{approxQuantile} and thus
proportionally reduce the size of the $\Delta k$ closest
returned to the driver in the last step of GK Select.

\newcolumntype{L}{>{\raggedright\arraybackslash}X} % left‐aligned X
\newcolumntype{M}{>{\centering\arraybackslash}X}   % centered X
\newcolumntype{B}{>{\raggedright\arraybackslash}p{0.2\textwidth}} % wide col
\newcolumntype{U}{>{\raggedright\arraybackslash}p{0.35\textwidth}} % ultrawide col

\begin{table*}[t]
\centering
\caption{Asymptotic executor and driver complexity for quantile methods.}
\label{tab:comparison-time-mem}
\scriptsize
\renewcommand{\arraystretch}{1.2}
\setlength{\tabcolsep}{4pt}
\begin{tabularx}{\textwidth}{lBMMMM}
\toprule
\textbf{Algorithm} & \textbf{Executor time} & \textbf{Driver time} & \textbf{Executor memory} & \textbf{Driver memory} & \textbf{Executor Work} \\
\midrule
Spark Full Sort &
$O\!\left(\tfrac{n}{P}\log\tfrac{n}{P}\right)$ &
$O(r P\log(r P))$ &
$O(\tfrac{n}{P})$ &
$O(r P)$ &
$O(n \log\tfrac{n}{P})$ \\

AFS &
$O\!\left(\tfrac{n}{P}\right)$ &
$O(\log n)$ &
$O(\tfrac{n}{P})$ &
$O(1)$ &
$O(n)$ \\

Jeffers Select &
$O\!\left(\tfrac{n}{P}\right)$ &
$O(P\log n)$ &
$O(\tfrac{n}{P})$ &
$O(P)$ &
$O(n)$ \\

Classical GK Sketch &
$O\!\left(\tfrac{n}{P}\log\tfrac{1}{\varepsilon} +
  \tfrac{n}{P}\log\log(\varepsilon\tfrac{n}{P})\right)$ &
n/a (streaming) &
$O\!\left(\tfrac{1}{\varepsilon}\log(\varepsilon\tfrac{n}{P})\right)$ &
n/a (streaming) &
$O\!\left(n\log\tfrac{1}{\varepsilon} +
  n \log\log(\varepsilon\tfrac{n}{P})\right)$ \\

Spark GK Sketch &
$O\!\left(\tfrac{n}{P}\log B +
  \tfrac{1}{\varepsilon}\tfrac{n}{P}\tfrac{1}{B}\log(\varepsilon\tfrac{n}{P})\right)$ &
$\Theta\!\left(\tfrac{P}{\varepsilon}\log(\varepsilon n)\right)$ &
$O\!\left(B + \tfrac{1}{\varepsilon}\log(\varepsilon\tfrac{n}{P})\right)$ &
$O\!\left(\tfrac{P}{\varepsilon}\log(\varepsilon\tfrac{n}{P})\right)$ &
$O\!\left(n \log B +
  \tfrac{1}{\varepsilon}\tfrac{n}{P}\tfrac{1}{B}\log(\varepsilon\tfrac{n}{P})\right)$ \\

GK Select &
$O\!\left(\tfrac{n}{P}\log\tfrac{1}{\varepsilon} + \tfrac{n}{P}\log\log(\varepsilon\tfrac{n}{P})\right)$ &
$O\left(\tfrac{P}{\varepsilon}\log(\varepsilon \tfrac{n}{P}) 
                    + \varepsilon n\right)$ &
$O\left(\tfrac{n}{P}\right)$ &
$O\left(\tfrac{P}{\varepsilon} \log(\varepsilon \tfrac{n}{P}) + \varepsilon n\right)$ &
$O\!\left(n\log\tfrac{1}{\varepsilon} + n \log\log(\varepsilon\tfrac{n}{P})\right)$  \\
\bottomrule
\end{tabularx}
\end{table*}

\begin{table}[t]
\centering
\caption{Communication and synchronization complexity for quantile methods.}
\label{tab:comparison-comm}
\scriptsize
\renewcommand{\arraystretch}{1.2}
\setlength{\tabcolsep}{4pt}
\begin{tabularx}{\columnwidth}{lMMMMM}
\toprule
\textbf{Algorithm} & \textbf{Network volume} & \textbf{Full Shuffles} & \textbf{Rounds} & \textbf{Persists} & \textbf{E/A} \\
\midrule
Spark Full Sort &
$O(n)$ &
1 & 1$^\dagger$ & 0 & Exact \\

AFS &
$O(P\log n)$ &
0 & $O(\log n)$ & $O(\log n)$ & Exact \\

Jeffers Select &
$O(P\log n)$ &
0 & $O(\log n)$ & $O(\log n)$ & Exact \\

Spark GK Sketch &
$O\!\left(\tfrac{P}{\varepsilon}\log(\varepsilon\tfrac{n}{P})\right)$ &
0 & 1 & 0 & Approx. \\

GK Select &
$O\!\big(\tfrac{P}{\varepsilon}\log(\varepsilon\tfrac{n}{P})$  & 0 & 3 & 0$^\ddagger$ & Exact \\
 & $+ \varepsilon n P\big)$ \\
\bottomrule
\end{tabularx}

\vspace{2pt}
{\raggedright
  
  $\dagger$\, In the conference version we presented this as two actions.  The sort has at least a stage boundary after the operations before the shuffle and then a stage boundary after the shuffle.  However, \texttt{orderBy} itself performs only one round.
\par 
  
$\ddagger$\,GK Select \emph{does not} call \texttt{persist}; the conference version incorrectly listed one persist.  
\par}
\end{table}

\section{Results}\label{Results}

\subsection{Experimental Setup}

In order to effectively compare the performance of the proposed GK
select algorithm, the execution time was measured across the following
data set sizes ranging from 1 million to 1 billion random
integers. The number of core nodes was varied across each of the data set
sizes, ranging from 12 partitions (3 core nodes) to 120 partitions (30
core nodes). Each trial was run 100 times, computing the mean time at the
end. The resulting times were compared across several algorithms: Full
sort, AFS, Jeffers, and GK Sketch. The GK Sketch algorithm in this
case is the Apache Spark implementation, \texttt{approxQuantile}.  We
used AWS EMR release emr-7.9.0 with instance type \texttt{m5.xlarge},
having 4 vCore, 16 GiB memory running Spark version 3.5.5 and Hadoop
version 3.4.1. Storage type is EBS using 15GiB volumes.

%\begin{figure}[H] 
%    \centering % Centers the image
%    \includegraphics[width=0.4\textwidth]{3_core_nodes.png} 
%    \caption{Performance with 3 core nodes.}
%
%    \label{3core} 
%\end{figure}
%
%In the three-core-node configuration shown in Figure 1, trends mirror the
%single-core results. Parallelism reduces absolute runtimes for all
%methods, but the relative ordering is unchanged: GK Sketch and GK
%Select achieve substantially lower runtimes than Jeffers, Al-Furaih,
%and full sort.

\begin{figure}[H] 
    \centering % Centers the image
    \includegraphics[width=0.4\textwidth]{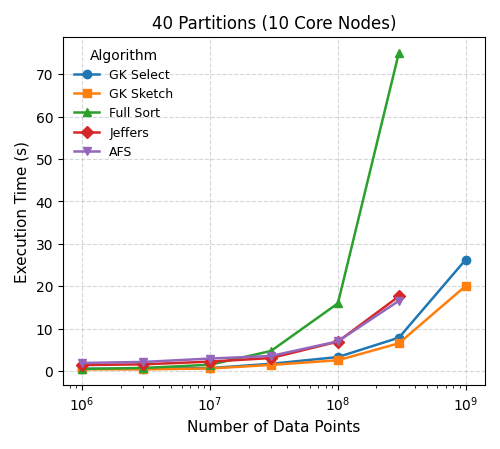} 
    \caption{Performance with 10 core nodes.}

    \label{10core} 
\end{figure}

\begin{figure}[H] 
    \centering % Centers the image
    \includegraphics[width=0.4\textwidth]{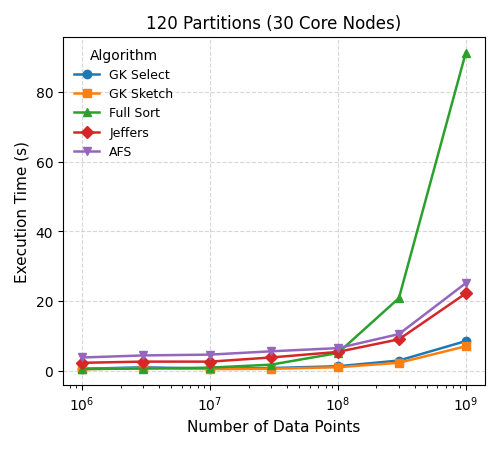} 
    \caption{Performance with 30 core nodes.}

    \label{30core} 
\end{figure}

In the 30-core-node configuration (Figure \ref{30core}), parallelism
compresses absolute runtimes across the board while preserving the
relative ordering observed at smaller core counts. GK Sketch and GK
Select remain dominant: they stay sub-second through $10^7$ points,
are only ~$1-1.3$ s at $10^8$, and increase to roughly $7-9$ s even at
$10^{9}$ points. This shallow growth indicates near-linear scaling
with small constant factors. Notably, these results show that even as
$n$ becomes very large, GK Select achieves exact computation with only
a modest time penalty—there is little practical sacrifice in runtime
for exactness under multi-core execution. By contrast, Full Sort
benefits from parallelism at moderate sizes but exhibits the expected
superlinear growth as $n$ rises, and Jeffers/Al-Furaih do not extend to
the largest inputs in this plot, suggesting less favorable scaling or
resource limits relative to the GK methods.

\subsection{Robustness across Data Distributions}                                               
                                         
To evaluate runtime stability across varied input characteristics, we
tested GK Select under four data distributions.

\textbf{Uniform}: values drawn independently and uniformly at random
from $[-10^9, 10^9)$; the baseline case with no skew or structure.
  
\textbf{Zipf}: values drawn from a Zipf distribution with exponent
$s=2.5$, mapped into $[-10^9, 10^9)$; a small number of values occur
with high frequency, modelling power-law data such as word
frequencies or access logs.

\textbf{Bimodal}: a 50/50 mixture of two Gaussian distributions
centered at $-3.33 \times 10^8$ and $+3.33 \times 10^8$, each with
standard deviation $1.66 \times 10^8$, clamped to $[-10^9, 10^9)$.
  
\textbf{Sorted}: each partition draws values uniformly at random from
a non-overlapping subrange of $[-10^9, 10^9)$ and sorts them locally,
  producing globally ordered data with each partition holding a
  contiguous band of values.

\begin{figure}[H]
    \centering
    \includegraphics[width=0.48\textwidth]{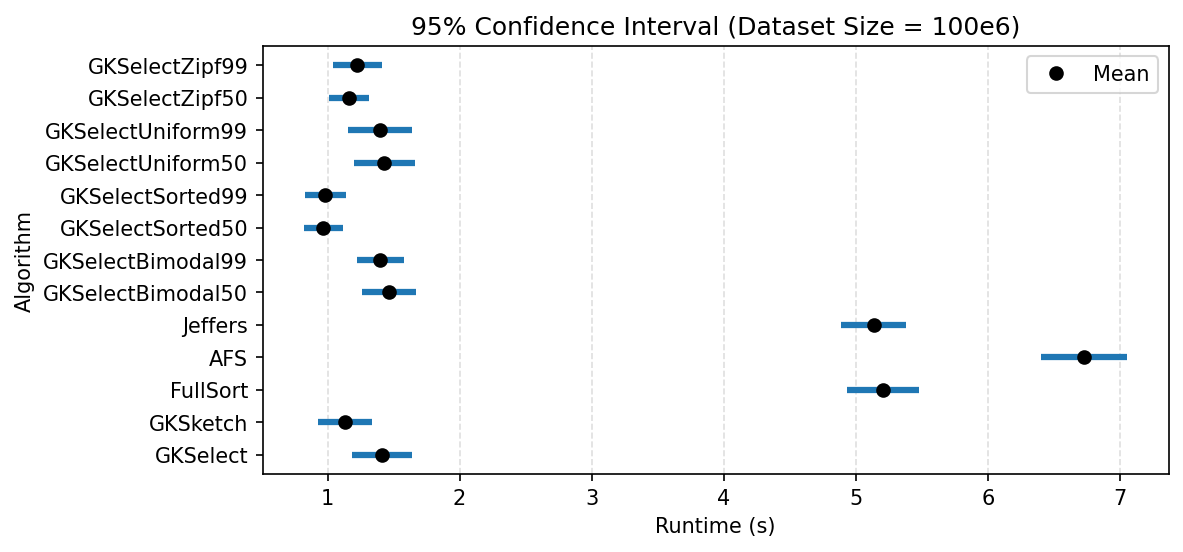}
    \caption{Runtime 95\% confidence intervals for data size $10^8$.
             The black circle marks the mean over 100 runs.
             Algorithm labels ending in \texttt{50} or \texttt{99}
             denote the 50th and 99th percentile targets, respectively.}
    \label{fig:CI-1e8}
\end{figure}

\begin{figure}[H]
    \centering
    \includegraphics[width=0.48\textwidth]{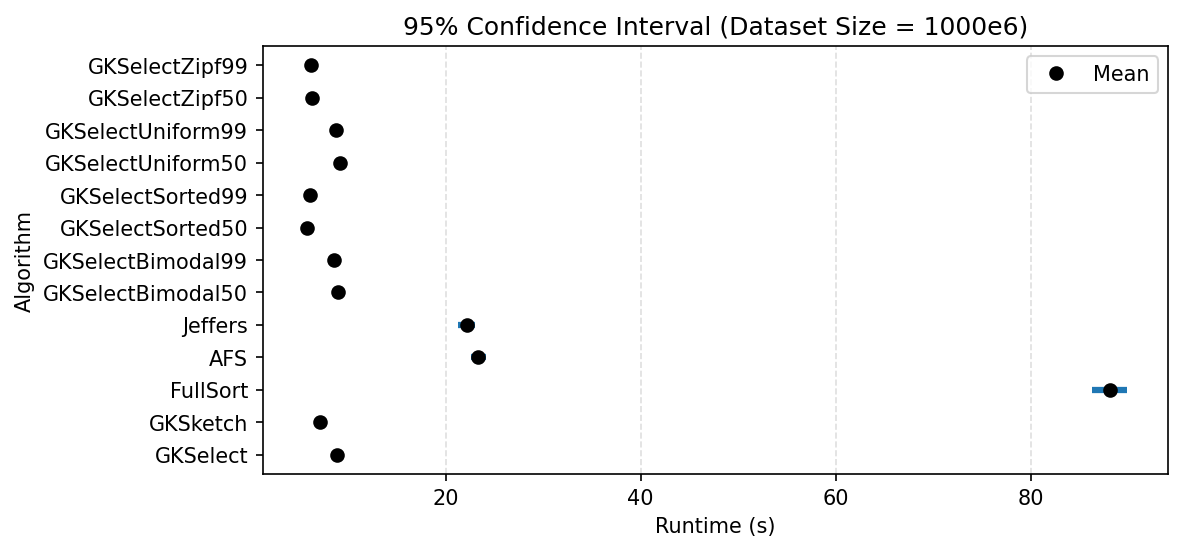}
    \caption{Runtime 95\% confidence intervals for data size $10^9$.
             The black circle marks the mean over 100 runs.
             Algorithm labels ending in \texttt{50} or \texttt{99}
             denote the 50th and 99th percentile targets, respectively.}
    \label{fig:CI-1e9}
\end{figure}

Execution times were measured over 100 runs and 95\% confidence
intervals computed using the $t$-distribution.
Figures~\ref{fig:CI-1e8}--\ref{fig:CI-1e9} show the results.

The confidence intervals are narrow and consistent across all four
distributions at both percentiles, including the highly skewed Zipf
distribution. This indicates that GK Select's runtime is not
meaningfully sensitive to the shape of the input distribution.

\section{Conclusion}\label{sec:conclusion}

We introduced \emph{GK Select}, an exact quantile algorithm that uses
a GK sketch–derived pivot to guide a Dutch partition.  By replacing
iterative count-and-discard rounds with a constant number of rounds,
GK Select reduces execution time while retaining exactness. Empirically,
it nearly matches the latency of Spark’s approximate GK operator and
outperforms full sort by up to $10\times$ across datasets to $10^9$
elements on 30 core nodes. These results demonstrate that exact
quantiles can be computed in Spark at near-approximate cost through
high-quality pivots and bounding the number of rounds.

\bibliographystyle{IEEEtran}
\bibliography{quantile}

\appendix

\section{GK Select: Scala/Spark Implementation}\label{app:gkselect}

The listing in Figure~\ref{lst:gkselect} is the core of
\texttt{GKSelectQuantile.scala} stripped of timing instrumentation,
logging, data generation, and verification scaffolding.

\begin{figure*}[t]
\begin{lstlisting}
def firstPass(it: Iterator[Int], pivot: Int): Iterator[(Long,Long,Long)] = {
  var lt, eq, gt = 0L
  it.foreach { v =>
    if      (v < pivot) lt += 1
    else if (v > pivot) gt += 1
    else                eq += 1
  }
  Iterator((lt, eq, gt))
}

def quickSelect(a: Array[Int], lo: Int, hi: Int, k: Int): Unit = {
  var (l, h) = (lo, hi)
  while (l <= h) {
    val pIdx = l + Random.nextInt(h - l + 1)
    val pVal = a(pIdx); a(pIdx) = a(h); a(h) = pVal
    var s = l; var i = l
    while (i < h) { if (a(i) < pVal) { val t=a(i); a(i)=a(s); a(s)=t; s+=1 }; i+=1 }
    val t = a(s); a(s) = a(h); a(h) = t
    if (s == k) return else if (s < k) l = s+1 else h = s-1
  }
}

def secondPass(it: Iterator[Int], pivot: Int, delta: Int): Iterator[Array[Int]] = {
  val a = it.toArray
  if (delta < 0) {                    // true rank is left of pivot
    var l = 0; var m = a.length-1; var r = a.length-1
    while (m >= l) {
      if      (a(m) > pivot) { val t=a(m); a(m)=a(r); a(r)=t; r-=1; m-=1 }
      else if (a(m) < pivot) { val t=a(m); a(m)=a(l); a(l)=t; l+=1        }
      else m -= 1
    }
    val tgt = math.max(0, l + delta)
    quickSelect(a, 0, l-1, tgt)
    Iterator(a.slice(tgt, l))
  } else {                            // true rank is right of pivot
    var l = 0; var m = 0; var r = a.length-1
    while (m <= r) {
      if      (a(m) < pivot) { val t=a(m); a(m)=a(l); a(l)=t; l+=1; m+=1 }
      else if (a(m) > pivot) { val t=a(m); a(m)=a(r); a(r)=t; r-=1        }
      else m += 1
    }
    val tgt = math.min(a.length-1, r + delta)
    quickSelect(a, r+1, a.length-1, tgt)
    Iterator(a.slice(r+1, tgt+1))
  }
}

def reduceSlices(a: Array[Int], b: Array[Int], delta: Int): Array[Int] = {
  val c = a ++ b
  if (c.length <= math.abs(delta)) c
  else if (delta < 0) { val tgt = c.length + delta
                        quickSelect(c, 0, c.length-1, tgt); c.slice(tgt, c.length) }
  else                { quickSelect(c, 0, c.length-1, delta); c.slice(0, delta) }
}

// --------- main algorithm (per quantile query) ----------------
val approxPivot = rdd.toDF("value")
  .stat.approxQuantile("value", Array(targetQ), epsilon).head.toInt   // Round 1
val (lt, eq, gt) =
  rdd.mapPartitions(firstPass(_, approxPivot))
     .reduce((a,b) => (a._1+b._1, a._2+b._2, a._3+b._3))            // Round 2
if (lt <= trueRank && trueRank < lt + eq) {
  approxPivot                                            // pivot is the exact answer
} else {
  val approxRank = if (lt + eq <= trueRank) lt + eq - 1 else lt
  val delta      = (trueRank - approxRank).toInt
  val finalSlice = rdd.mapPartitions(secondPass(_, approxPivot, delta))
    .treeReduce((a,b) => reduceSlices(a, b, delta), depth = depth)    // Round 3
  if (delta < 0) finalSlice.min else finalSlice.max
}
\end{lstlisting}
\caption{GK Select core (Scala/Spark).  \texttt{delta} is the signed
  offset from \texttt{approxPivot}'s rank to \texttt{trueRank}; its
  sign indicates which side of the pivot contains the target element.
  \texttt{secondPass} performs a Dutch partition on each executor's
  local array and returns only the $|\mathtt{delta}|$-element boundary
  slice.  \texttt{reduceSlices} combines two slices during
  \texttt{treeReduce}, discarding elements that cannot be the answer.}
\label{lst:gkselect}
\end{figure*}

\end{document}